\documentstyle[11pt,paspconf]{article}

\begin{document}

\title{ A magnetic model for acoustic modes in roAp stars }
\author{E.J. Zita}
\affil{ The Evergreen State College, Olympia, WA, 98505}

\begin{abstract}
The mechanism for excitation of p-modes in rapidly oscillating, peculiar A (roAp, or cool chemically peculiar, CP) stars is unknown.  Observations strongly suggest that acoustic modes in roAp stars are causally linked to the stars' magnetic field.  We propose that small fluctuations in the shape of the mean magnetic field drive magnetosonic waves, which are observed as p-modes in these stars.
	The dynamic edge region of roAp stars is a force-free spherical shell.  When strongly coupled to the magnetic field, a force-free plasma can oscillate about a minimum in its mean magnetic energy.  We describe the stable eigenmodes for this energy minimum in a spherical shell with an open boundary.  The wavenumbers, frequencies, and energies of resulting oscillations are consistent with observations of p-modes in roAp stars.  Our magnetic model for p-mode oscillations in stars does not invoke convection or opacity mechanisms.  We also suggest the possibility of a nonlinear dynamo for such magnetic stars, which lack the convection needed for the usual (-( dynamo.
\end{abstract}


\section{Introduction}

	Magnetic, rapidly oscillating, peculiar A stars ring with sound waves (p-modes).  Fourier transforms of millimagnitude luminosity fluctuations reveal unusually robust, evenly-spaced millihertz peaks, characteristic of p-modes in the edge of the stars.  Oscillating stars tend to have strong (kG) constant magnetic fields which appear strongly coupled to the stellar dynamics.  Variations in apparent mean field strength are consistent with the oblique pulsator model, in which symmetry axes of the magnetic field and of the stellar pulsations are aligned with each other, and are inclined with respect to the rotation axis (Kurtz, 1990; and Unno et al., 1989).  

	The cause of the observed nonradial oscillations and the role of the magnetic field in these stars is unknown.  While convection above the radiation zone of the sun drives p-modes in the photosphere, roAp stars have little or no convection near the surface.  (Their temperature gradients are too low to require convective transport, and their high surface abundances of metals preclude convective mixing.)  While opacity mechanisms can drive oscillations in Cepheids, partial ionization zones in roAp stars are radiatively damped (Dziembowski and Goode, 1996).  (If surface ionization zones did survive in roAp stars, they should also drive p-modes in nonmagnetic stars, which are not detected.)  While magnetic fields clearly perturb oscillations in stars such as white dwarfs, perturbation approaches are of limited applicability in roAp stars, where magnetic energy is not much smaller than gravitational and thermal energies in the dynamic edge region (Shibahashi and Takata, 1993).  A new approach may shed light on the source of these p-modes.

	We propose a resonant, magnetic driver for acoustic modes in roAp stars.  We start with the general characteristics of a force-free spherical shell of plasma (Section 2).  Oscillations about these magnetic energy minima can yield magnetosonic waves (Section 3).  We compare calculations from our model to observations of roAp stars (Section 4).  Finally, we  summarize and discuss future work (Section 5).

\section{ Magnetic energy minima in force-free spherical shells}

	Magnetic plasma is "force-free" insofar as pressure is uniform, that is over distances much less than the local pressure scale height.  In such regions, the equation of motion reduces to JxB=0, and plasma currents are parallel to their magnetic fields (LŸst and SchlŸter 1954).  Many stars are force-free in their outer regions, which tend to contain little mass and to radiate much of the stars' observed light.  Only those force-free regions with magnetic pressure comparable to gas pressure () will also have plasma motions strongly coupled to the magnetic field.

	Force free fields satisfy , where the parallel current parameter  is proportional to the helicity (or twist) of the magnetic field (Chandraskehar and Kendall 1957).  Further, constant-( fields minimize magnetic energy (Woltjer 1958); this insight is key to our model.  In spherical geometries, nontrivial (((0) force-free fields are intrinsically twistier than dipole, with stronger multipole moments for higher parallel current.  Spherical fields are expressed in terms of a poloidal function which satisfies the scalar Helmholtz equation , where the radial wavefunctions plm(r) are spherical Bessel functions (Berger, 1985).  The solution B=curl2 x P(x) + curl x T(x) depends on the geometry and boundary conditions of the force-free plasma.  

	The solution for an open outer boundary at outer radius R2 (and a dipole inner boundary at inner radius R1) is found by conserving relative helicity in a force-free spherical shell (Berger 1985).  The resultant magnetic energy density W is minimized for unique values of the parallel current parameter (.  For example, W(() for a roAp special case (Sec.3) is shown in Fig.1.  Only the first nontrivial magnetic energy minimum, at (0, is stable to perturbations, including pressure perturbations, other helicity-conserving perturbations, and even small helicity non-conserving perturbations.

	By calculating Berger's magnetic energy minimum (MEM) for a series of spherical shells with variable thickness ÆR = R2-R1/R1, we also find that the energy density and (0 increase nonlinearly for thinner shells.  This means that, for a given surface field strength, a thinner force-free region has a twistier magnetic equilibrium, with proportionally more energy available to drive plasma waves.

\section{ Magnetic model for oscillations }

	For a given shell thickness ÆR, the MEM field has a unique parallel current parameter (0, which describes an equilibrium twist for the field.  Consider a dipole flux tube slightly twisted by higher multipoles.  The shape of the force-free field can oscillate about its equilibrium configuration.  At higher (+, the field is twisted (or compressed) more by greater parallel current.  At lower (-, reduced parallel current over-relaxes (or stretches) the twisted field.  The magnetic energy stored and released by these oscillations corresponds to the excursions from (0 in the W(() well.  The magnetic fluctuations can be visualized as a spring compressing and stretching, as the current parameter increases and decreases.  These changes in helicity of the force-free field are longitudinal oscillations about the MEM state.

	Let the magnetic field be written B=B0 + B1, where B0 is the mean field and B1 is the fluctuating part of the field.  Consider a region where B0(B( , as in a primarily dipole field.  The longitudinal (B0B1) oscillations about the MEM generate a circulating electric field E1, by Faraday's law, as in Fig.2a.  (E0=0 in the plasma frame, as usual.)  This drives a vr=E1B1 drift of radial compressions, which is simply a magnetosonic wave (propagating both inward and outward from the force-free region) resonating with the field fluctuations.  The oscillations in plasma pressure which constitute these magnetosonic waves may be observed as surface p-modes.  The geometry of these resonant modes acoustic modes will naturally reflect the global structure of the mean field.

	Recall that this "fast" magnetosonic mode travels across magnetic field lines (in distinction to the "slow" AlfvŽn wave, which plucks field lines and travels parallel to the field).  Magnetic modes can be a combination of AlfvŽnic and magnetosonic, depending on their direction of propagation with respect to the field lines.  Insofar as the natural oscillations in the shape of our force-free fields are longitudinal, perturbations to the mean field will preferentially excite magnetosonic waves.  

	This model naturally leads to a nonlinear dynamo which may sustain the star's magnetic field as its oscillations transfer energy to p-modes.  Even if the volume averages of the radial velocity fluctuations <vr> and of the longitudinal magnetic field <B1> fluctuations vanish, their cross-product can have a non-zero volume average <vrxB1> (Moffatt 1978).  The resultant azimuthal electric field E( can continue driving the mean magnetic field B0, as in Fig.2b.  Such a nonlinear, self-sustaining process has been shown to drive the dynamo of force-free, periodic-cylindrical Reversed Field Pinch plasmas (Schnack et al. 1985; Terry et al. 1991), due to the helical eigenmodes about the magnetic energy minimum of this system (Taylor 1979).

\section{ Compare model calculations and roAp observations }

	The force-free spherical shell at the edge of roAp stars coincides with the dynamic region for acoustic modes in the stars (Zita 1996).  The field strength and the hydrostatic equilibrium of the star directly determine the size ÆR and parallel current parameter ( of the force-free shell, as described in Section 2.  Taken together, these yield values for the wavelength, magnetosonic frequency, and energy of the oscillations.
	Consider a conservative case:  a thick force-free shell (with correspondingly low magnetic energy density) that is barely coupled to a relatively hot plasma.  Recall that thinner-shell MEM fields will have greater magnetic energy available to drive oscillations.  Cooler plasmas (with lower gas pressures) will be more strongly coupled to their magnetic fields.
	We calculate the size of the force-free region from a hydrostatic equilibrium model (Kawaler 1994) for a roAp star with M=3M( and R=1.92 R(.  We find that the pressure gradient is low in the outer few percent of the star.  Plasma-field coupling becomes significant where the ratio of gas pressure to magnetic pressure  approaches 1 or less, in the dynamic edge region (Zita 1996).

	Magnetic energy:
	Assuming a force-free shell of thickness ÆR=0.05, we use Berger's method to calculate the solutions shown in Fig.1.  We find the nontrivial magnetic energy minimum at a parallel current parameter of (0~90 with magnetic energy density W=212.  The magnetic field is normalized to the mean field strength B0 and lengths are measured in units of the inner radius R1.  For a mean field of B0 =104 G in Kawaler's hydrostatic equilibrium, this yields a magnetic energy of over 1042 ergs.
	Compare this to the kinetic energy of the plasma oscillations.  The p-mode energy is overestimated by assuming the entire dynamic shell (of density 7x10-7 g/cc) moves with the maximum observed speed (of 20,000 cm/s) (Matthews 1997).  These acoustic modes require about 1035 ergs of kinetic energy.  As little as 10-6 of the mean magnetic energy could drive the p-modes, in this case.
	The fraction of the mean magnetic energy actually available to acoustic oscillations depends in part on the amplitude of the magnetic excursions from the equilibrium (0.  Our example requires fluctuations on the order of 10 mG about a mean field strength of 104 G.  Resistive and AlfvŽnic losses could necessitate higher fluctuations.  Such small magnetic oscillations are consistent with observations:  they are below detection limits (Kurtz 1997).

	Radial wavelength:  
	The parallel current parameter ( is inversely proportional to the length scale on which the mean magnetic field twists, and proportional to the number of radial nodes.  We plot the radial wavefunctions plm(r) of the force-free field to see the location of radial nodes (Fig.3).  For the ÆR=0.05 case, inspection of the radial wavefunction reveals a wavelength of about 6

	Magnetosonic frequencies:
	The magnetosonic modes described in Section 3 have frequencies ( given by , where the sound speed , the AlfvŽn speed , and the wavenumber k=(/R1.  In our example, the magnetosonic frequency is independent of field strength except at very high fields, since if vS>>vA, then (~ vS k.   For fields up to 10 kG, calculated frequencies are about 3 mHz and measured fundamental frequencies cluster around 1-3 mHz (Kurtz 1990).  Frequencies will depend more strongly on field strength in cooler stars.  Since frequency calculations also depend strongly on the hydrostatic equilibrium, estimates of the size of each observed star will be important for more detailed comparisons with observations.

\section{ Summary and future work }

We have described a magnetic model that addresses theoretical and observational questions:
What is the physical nature of oscillations about a spherical-shell magnetic energy minimum (MEM) configuration with an open boundary?
What causes the observed p-modes in roAp stars, if not convection or opacity mechanisms?
Why do these modes have the same global structure and symmetry axis as the magnetic field?

Since the dynamic edge plasma in roAp stars is force-free and low-(, oscillations about the minimum energy configuration of the magnetic field are to be expected.  Longitudinal magnetic eigenmodes  can drive magnetosonic waves which are detected as surface p-modes.  Combining a force-free field solution with a hydrostatic equilibrium yields wavelengths, frequencies, and energies for a given field strength.  Calculated results are consistent with observations.  The magnetic energy minimum of a force-free shell provides a stable well and adequate energy to support p-mode oscillations in roAp stars.
	Our magnetic model can be further tested with a set of hydrostatic equilibria spanning the roAp class.  Given the size (R of the force-free shells, a corresponding set of MEM can be calculated, yielding oscillation spectra and energies.  In addition, the force-free fields can be decomposed into their multipole moments.  This library of predictions can be compared to subsequent observations of the sizes, temperatures, oscillation spectra, surface metal distibution, and magnetic field strength and structure of roAp stars.  Numerical simulations can characterize in more detail the turbulent dynamo sketched here.

\acknowledgments

This work has been supported by the National Science Foundation, Grinnell College, and Iowa State University.

We gratefully acknowledge helpful discussions with Steve Kawaler and Lee Anne Willson (Ames, Iowa),  Mitchell Berger (Cambridge),  Don Kurtz (Cape Town),  Hiromoto Shibahashi (Tokyo), Wojciech Dziembowski (Warszawa),  S¿ren Frandsen and J¿rgen Christensen-Dalsgaard (Aarhus),  Jaymie Matthews (Victoria, BC), and others.

\begin{question}
Does your model imply that nonmagnetic Ap stars will not exhibit p-modes?
\end{question}
\begin{answer}{Zita}
Yes;  if nonmagnetic Ap stars did show p-modes, they could not be due to the mechanism proposed here.
\end{answer}

\begin{question}
Why wouldn't hotter (O or B) magnetic stars also show p-modes?
\end{question}
\begin{answer}{Zita}
The plasma is too weakly coupled to 10 kG or weaker magnetic fields in hotter stars, where.
\end{answer}

\begin{question}
What drives the magnetic oscillations?
\end{question}
\begin{answer}{Zita}
The mean field can be perturbed near its equilibrium state by thermal motions, turbulence, fluid flow, pressure perturbations, or other processes.  As long as the perturbations transport little energy or helicity, and have little length coherence, the self-forces of the perturbed field will relax it back toward the minimum of its W(() well.
\end{answer}

\begin{question}
What is the relationship between the chemical peculiarities, the strong magnetic field, and the pulsations in roAp stars?
\end{question}
\begin{answer}{Zita}
(Nontrivial) force-free plasmas have not just dipole, but higher multipole magnetic fields (in a spherical shell).  If metals preferentially drift to the surface along field lines, the patchy distributions found by Rainer Kuschnig et al. may provide an independent test of field structures. Since our model predicts p-modes of the same global structure as the mean field, these fields could also excite l(2 modes.
\end{answer}

\end{document}